\documentstyle[12pt]{article}
\begin{document}

\title{\Large\bf On the equivalence among some chiral-boson theories}

\author{R. Amorim\thanks{\noindent Electronic mails: ift01001 @ ufrj
and amorim @ if.ufrj.br} ~and J. Barcelos-Neto\thanks{\noindent
Electronic mails: ift03001 @ ufrj and barcelos @ vms1.nce.ufrj.br}\\
Instituto de F\'{\i}sica\\ Universidade Federal do Rio de Janeiro\\
RJ 21945-970 - Caixa Postal 68528 - Brasil}
\date{}
\maketitle
\abstract
We make a comparative study of chiral-boson theories in the
Florenani-Jackiw (FJ) and linear constraint formulations. A special
attention is given to the case with an improved way of implementing
the linear constraint. We show that it has the same spectrum of the
FJ formulation.

\vfill
\noindent PACS: 03.70.+k, 11.10.Ef, 11.15.-q
\vspace{1cm}
\newpage

\section{Introduction}

Chiral-boson theories are two-dimensional scalar theories where the
scalar field $\phi$ exhibits the so-called chiral condition
$\dot\phi=\phi^\prime$ as a solution
\footnote{Dot means time derivative and prime means derivative with
respect to the spatial coordinate x. We shall also consider the
metric convention $\eta_{00}=-\eta_{11}=-1$, $\eta_{01}=0$ throughout
this paper.}.
It is usually known four formulations of chiral-boson theories in
literature. The one that has called much attention is due to
Floreanini and Jackiw (FJ)~\cite{FJ}, described by the non-covariant
Lagrangian

\begin{equation}
{\cal L}_1=\dot\phi\phi^\prime-\phi^\prime\phi^\prime\,.
\label{1.1}
\end{equation}

\bigskip\noindent
This theory has a constraint that is second-class at all points of
momentum space~\cite{Dirac,Gi1} except one, where it is
first-class~\cite{Gi2}.  This fact deserves some care when
calculating the spectrum, because raising and lowering operators have
to be defined at all points~\cite{Gi2}.

\medskip
Another formulation is due to Srivastava, where the chiral condition
is introduced linearly by means of a Lagrange multiplier~\cite{Prem}.
The corresponding Lagrangian reads

\begin{equation}
{\cal L}_2=-{1\over2}\,\eta_{\mu\nu}\,
\partial^\mu\phi\partial^\nu\phi
+\lambda\,\bigl(\dot\phi-\phi^\prime\bigr)\,.
\label{1.2}
\end{equation}

\bigskip\noindent
This theory has been criticized by some authors~\cite{Ha,Gi3}. The
main arguments are that it does not lead to a positive definite
Hamiltonian and that its physical spectrum, after introducing ghost
degrees of freedom, is just the vacuum state. In fact, the theory
described by~(\ref{1.2}) is not equivalent to the FJ one, even though
both of them contain the same chiral condition
$\dot\phi-\phi^\prime=0$ as a classical equation of motion. We
deserve Sec.~2 to discuss these points with details and to review the
obtainment of the spectrum of these two theories. We also take the
opportunity to introduce some new ingredients and to fix the notation
and convention we shall use throughout the paper.

\medskip
More recently, it has been introduced another way of implementing the
linear constraint in the chiral-boson theory~\cite{Ki,Ba}. This is
described by the Lagrangian

\begin{equation}
{\cal L}_3=-{1\over2}\,\eta_{\mu\nu}\,
\partial^\mu\phi\partial^\nu\phi
+\lambda\,\bigl(\dot\phi-\phi^\prime\bigr)
+{1\over2}\,\lambda^2\,.
\label{1.3}
\end{equation}

\bigskip\noindent
Here, the equation of motion for $\lambda$ does not lead to the
chiral condition, but using this equation back to~(\ref{1.3}) the FJ
Lagrangian is obtained. They are consequently on-shell classically
equivalent (as it was in some sense FJ and Srivastava formulations).
The main purpose of our work is to study the theory described by the
Lagrangian above in the quantum point of view. This will be done in
Sec.~3.  We shall see that it does not have the inconsistencies of
the previous formulation and, more than that, has the same spectrum
of the FJ theory. In other words, they are also quantically
equivalent.

\medskip
To conclude this introduction we just mention the formulation due to
Siegel~\cite{Si}, where the chiral condition is obtained by
projecting out one of the components of the energy-momentum tensor,
which results in a Lagrangian where the chiral relation appears
quadratically

\begin{equation}
{\cal L}_4=-{1\over2}\,\eta_{\mu\nu}\,
\partial^\mu\phi\partial^\nu\phi
+\lambda\,(\dot\phi-\phi^\prime)^2\,.
\label{1.4}
\end{equation}

\bigskip\noindent
This system is anomalous and its quantization is only consistently
achieved by introducing a Wess-Zumino term, resulting a theory where
chiral-bosons are coupled to gravity~\cite{Im}.

\bigskip
\section{Spectra of FJ and Srivastava chiral-boson theories}
\bigskip
The Lagrangian of the FJ theory leads to the constraint

\begin{equation}
\Omega_1(x,t)=\pi(x,t)-\phi^\prime(x,t)\,,
\label{2.1}
\end{equation}

\bigskip\noindent
where $\pi$ is the canonical momentum conjugate to $\phi$.  The
quantity $\Omega_1$ represents, in fact, an infinite number of
constraints, one for each value of the space coordinate $x$. Another
important particularity concerns with the constraint matrix

\begin{eqnarray}
C(x,y)&=&\bigl\{\Omega_1(x,t),\,\Omega_1(y,t)\bigr\}\,,
\nonumber\\
&=&-\,2\,\delta^\prime(x-y)\,.
\label{2.2}
\end{eqnarray}

\bigskip\noindent
Its inverse reads

\begin{equation}
C^{-1}(x,y)=-\,\frac{1}{2}\,\epsilon(x-y)+f(t)\,,
\label{2.3}
\end{equation}

\bigskip\noindent
where $\epsilon(x-y)$ is the step function and $f(t)$ is some
arbitrary function of time. We see that the inverse is not unique.
This lack of uniqueness is related to the fact that there is one
first-class constraint among all the infinite constraints represented
by $\Omega_1$~\cite{Dirac,Gi1}. This is easily seen by considering
Fourier transformations of fields and constraints. They are defined
through the relations

\begin{eqnarray}
\phi(x,t)&=&\frac{1}{\sqrt{2\pi}}\,\int dk\,
e^{ikx}\,\tilde\phi(k,t)\,,
\nonumber\\
\pi(x,t)&=&\frac{1}{\sqrt{2\pi}}\,\int dk\,
e^{ikx}\,\tilde\pi(k,t)\,,
\nonumber\\
\Omega_1(x,t)&=&\frac{1}{\sqrt{2\pi}}\,\int dk\,
e^{ikx}\,\tilde\Omega_1(k,t)\,.
\label{2.4}
\end{eqnarray}

\bigskip\noindent
Using the relations above and~(\ref{2.1}) we see that

\begin{equation}
\tilde\Omega_1(k,t)=\tilde\pi(k,t)-i\,k\,\tilde\phi(k,t)\,.
\label{2.5}
\end{equation}

\bigskip\noindent
Further, the fundamental Poisson bracket relation, written in
momentum space,

\begin{equation}
\bigl\{\tilde\phi(k,t),\,\tilde\pi(p,t)\bigr\}=\delta(k+p)\,,
\label{2.6}
\end{equation}

\bigskip\noindent
gives for the constraint matrix

\begin{equation}
\tilde C(k,p)=-\,2\,i\,k\,\delta(k+p)\,.
\label{2.7}
\end{equation}

\bigskip
We notice that

\begin{equation}
\tilde\Omega_1(0,t)\equiv\tilde\Omega_2=\tilde\pi(0,t)
\label{2.8}
\end{equation}

\bigskip\noindent
is first-class. This explains why the inverse $C^{-1}(x,y)$ given
by~(\ref{2.3}) is not unique.

\medskip
To discuss the spectrum of the FJ theory, we have to be aware of the
quantization for all $k$. Let us then choose a gauge-fixing condition
for the first-class constraint $\tilde\pi(0,t)\approx0$. A natural
choice is

\begin{equation}
\tilde\Omega_3=\tilde\phi(0,t)\approx0\,.
\label{2.9}
\end{equation}

\bigskip
Let us calculate the Dirac brackets in order to perform the
quantization. The matrix of the Poisson brackets of constraints reads

\begin{equation}
\tilde C(\bar k,\bar p)=\left(
\begin{array}{ccc}
-2i\bar k\,\delta(\bar k+\bar p)&0&0\\
0&0&-\delta(0)\\
0&\delta(0)&0
\end{array}
\right)\,.
\label{2.10}
\end{equation}

\bigskip\noindent
We are using the notation $\bar k$ to mean any $k\not=0$. Considering
the inverse of this matrix, the Dirac brackets can be directly
calculated. The result is

\begin{eqnarray}
\bigl\{\tilde\phi(\bar k,t),\,\tilde\phi(\bar p,t)\bigr\}_D
&=&\frac{i}{2\bar k}\,\delta(\bar k+\bar p)\,,
\nonumber\\
\bigl\{\tilde\phi(0,t),\,\tilde\phi(\bar p,t)\bigr\}_D
&=&0\,,
\nonumber\\
\bigl\{\tilde\phi(0,t),\,\tilde\phi(0,t)\bigr\}_D
&=&0\,.
\label{2.11}
\end{eqnarray}

\bigskip
Before replacing the above Dirac brackets by commutators, it is
convenient to have a better view of the meaning of $\tilde\phi(k,t)$.
First we notice that the Lagrangian~(\ref{1.1}) leads to the equation
of motion

\begin{equation}
\dot\phi^\prime-\phi^{\prime\prime}=0\,.
\label{2.12}
\end{equation}

\bigskip\noindent
Using the Fourier transformations~(\ref{2.4}), we get

\begin{equation}
k\,\bigl(i\,\dot{\tilde\phi}+k\,\tilde\phi\bigr)=0\,.
\label{2.13}
\end{equation}

\bigskip\noindent
For $k\not=0$ the solution is

\begin{equation}
\tilde\phi(\bar k,t)=A(\bar k)\,e^{i\bar kt}\,,
\label{2.14}
\end{equation}

\bigskip\noindent
where $A$ is some generic function of $\bar k$. For $k=0$, nothing
can be concluded from eq.~(\ref{2.13}). However, from the gauge
condition we have chosen, the field $\tilde\phi$ is also defined at
this point (see expression~\ref{2.9}). So, one can consider the
expression above for all $k$, once one takes

\begin{equation}
A(0)=0\,.
\label{2.15}
\end{equation}

\bigskip\noindent
Replacing back these results into~(\ref{2.11}), and defining
$A^\dagger(\bar k)\equiv A(-\bar k)\equiv a(\bar k)/\sqrt{2\bar k}$
($\bar k>0$), $a(0)=0$, we obtain

\begin{eqnarray}
\bigl[a(\bar k),\,a^\dagger(\bar p)\bigr]
&=&\delta(\bar k-\bar p)\,,
\nonumber\\
\bigl[a(0),\,a^\dagger(0)\bigr]
&=&0\,\,=\bigl[a(0),\,a^\dagger(\bar p)\bigr]
=\bigl[a(0),\,a(\bar p)\bigr]
\hskip.5cm{\rm etc.}
\label{2.19}
\end{eqnarray}

\bigskip\noindent
where Dirac brackets have been replaced by $i$ times the
corresponding commutators. it is easily seen that $a^\dagger(\bar k)$
and $a(\bar k)$ are raising and lowering operators respectively. One
can also write the Fourier expansion for $\phi(x,t)$ as

\begin{equation}
\phi(x,t)=\frac{1}{\sqrt{2\pi}}\,\int_0^\infty
\frac{dk}{\sqrt{2k}}\Bigl[
a(k)\,e^{-ik(x+t)}+a^\dagger(k)\,e^{ik(x+t)}\Bigr]\,.
\label{2.20}
\end{equation}

\bigskip\noindent
It is interesting to notice that $\phi(x,t)=\phi(x^+)$ with
$x^+=x+t$. This reflects the presence of the chiral condition
$\dot\phi-\phi^\prime=0$ in the solution above.

\medskip
Let us now pass to consider the Hamiltonian in terms of the operators
$a(k)$ and $a^\dagger(k)$. Since commutators are obtained from Dirac
brackets, the quantum evolution ought to take constraints in a strong
way.  Consequently, the Hamiltonian we have to consider is the
canonical one and we use the constraints, when necessary, as strong
relations. Hence

\begin{eqnarray}
H_c&=&\int dx\,\bigl(\pi\dot\phi-{\cal L}\bigr)\,,
\nonumber\\
&=&\int dx\,\bigl[\bigl(\pi-\phi^\prime\bigr)\,\dot\phi
+\phi^\prime\phi^\prime\bigr]\,,
\nonumber\\
&\rightarrow&\int dx\,\phi^\prime\phi^\prime\,,
\label{2.21}
\end{eqnarray}

\bigskip\noindent where the constraint~(\ref{2.1}) was used in the
last step. The combination of (\ref{2.20}) and (\ref{2.21}) gives

\begin{equation}
H_c=\int_0^\infty dk\,k\,a^\dagger(k)\,a(k)\,,
\label{2.22}
\end{equation}

\bigskip\noindent
rewritten in normal order. Since the Hamiltonian above is positively
defined, we have that the action of the lowering operator in a state
cannot be done indefinitely. So, we can introduce a vacuum state from
where all states can be generated. Consequently, the FJ chiral boson
theory has a spectrum of bosonic massless particles just with right
movers.

\bigskip
Let us now discuss on the spectrum of the theory described by
Lagrangian (\ref{1.2}). It does not describe the same theory of the
FJ one as it will become patent during the development we are going
to present. However, there is a simple argument we can mention that
makes the affirmative above quite evident. To see this, we refer to
the constraints of (\ref{1.2}):

\begin{eqnarray}
\Omega_1&=&q\approx0\,,
\label{2.23}\\
\Omega_2&=&\pi-\phi^\prime-\lambda\approx0\,,
\label{2.23a}
\end{eqnarray}

\bigskip\noindent
where $q$ is the canonical momentum conjugated to $\lambda$. These
are second-class. Consequently, the theory has two physical
(continuous) degrees of freedom whereas the FJ has just one.

\medskip
The nonvanishing brackets involving $\Omega_1$ and $\Omega_2$ are

\begin{eqnarray}
\bigl\{\Omega_1(x,t),\,\Omega_2(y,t)\bigr\}
&=&\delta(x-y)\,,
\nonumber\\
\bigl\{\Omega_2(x,t),\,\Omega_2(y,t)\bigr\}
&=&-\,2\,\delta^\prime(x-y)\,.
\label{2.24}
\end{eqnarray}

\bigskip\noindent
Also here, let us make Fourier transformations of fields and
constraints. We obtain

\begin{eqnarray}
\tilde\Omega_1(k,t)&=&\tilde q(k,t)\,,
\label{2.25}\\
\tilde\Omega_2(k,t)&=&\tilde\pi(k,t)
-i\,k\,\tilde\phi(k,t)-\tilde\lambda
\label{2.26}
\end{eqnarray}

\bigskip\noindent The corresponding Poisson brackets matrix reads

\begin{equation}
C(k,p)=\left(
\begin{array}{cc}
0&1\\
-1&-2ik
\end{array}
\right)
\,\delta(k+p)\,,
\label{2.27}
\end{equation}

\bigskip\noindent
and we notice that $\det C\not=0$, independently of $k$. There is no
point where constraints can be first-class. This is another
difference with respect the FJ theory. The inverse $C^{-1}$ is
immediately calculated

\begin{equation}
C^{-1}(k,p)=\left(
\begin{array}{cc}
-2ik&-1\\
1&0
\end{array}
\right)
\,\delta(k+p)\,.
\label{2.28}
\end{equation}

\bigskip\noindent
The Dirac brackets involving $\tilde\phi$ and $\tilde\lambda$ are

\begin{eqnarray}
\bigl\{\tilde\phi(k,t),\,\tilde\phi(p,t)\bigr\}_D&=&0\,,
\nonumber\\
\bigl\{\tilde\lambda(k,t),\,\tilde\lambda(p,t)\bigr\}_D
&=&2\,i\,k\,\delta(k+p)\,,
\nonumber\\
\bigl\{\tilde\phi(k,t),\,\tilde\lambda(p,t)\bigr\}_D
&=&\delta(k+p)\,.
\label{2.29}
\end{eqnarray}

\bigskip
As it was done in the FJ case, let us look at equations of motion
before quantizing this theory. From Lagrangian~(\ref{1.2}) we get

\begin{eqnarray}
\Box\phi-\dot\lambda+\lambda^\prime&=&0\,,
\label{2.30}\\
\dot\phi-\phi^\prime&=&0\,.
\label{2.31}
\end{eqnarray}

\bigskip\noindent
The combination of these two equations gives

\begin{equation}
\dot\lambda-\lambda^\prime=0\,,
\label{2.32}
\end{equation}

\bigskip\noindent
that is, $\lambda$ also satisfies the chiral condition. Considering
the Fourier transformations for $\phi$ and $\lambda$ into
(\ref{2.31}) and (\ref{2.32}) we may rewrite  $\tilde\phi$ and
$\tilde\lambda$ by

\begin{eqnarray}
\tilde\phi(k,t)&=&A^\dagger(k)\,e^{ikt}\,,
\nonumber\\
\tilde\lambda(k,t)&=&\Lambda^\dagger(k)\,e^{ikt}\,.
\label{2.33}
\end{eqnarray}

\bigskip\noindent
With these results, the initial expressions of Fourier transformations
turns to be

\begin{eqnarray}
\phi(x,t)=\frac{1}{\sqrt{2\pi}}\,\int_0^\infty dk\,
\Bigl[A(k)\,e^{-ik(x+t)}
+A^\dagger(k)\,e^{ik(x+t)}\Bigr]\,,
\nonumber\\
\lambda(x,t)=\frac{1}{\sqrt{2\pi}}\,\int_0^\infty dk\,
\Bigl[\Lambda(k)\,e^{-ik(x+t)}
+\Lambda^\dagger(k)\,e^{ik(x+t)}\Bigr]\,.
\label{2.34}
\end{eqnarray}

\bigskip\noindent
The presence of the on-shell chiral conditions for both $\phi$ and
$\lambda$ is expressed in their dependence of $x^+=x+t$.

\medskip
Let us now consider the canonical Hamiltonian. The final result is
(using constraints strongly)

\begin{equation}
H_c=\int dx\,\phi^\prime\,\bigl(\phi^\prime+\lambda\bigr)\,,
\label{2.35}
\end{equation}

\bigskip\noindent
In terms of operators $A$, $A^\dagger$, $\Lambda$ and
$\Lambda^\dagger$ we get (in the normal order)

\begin{equation}
H_c=\int_0^\infty dk\,\Bigl[2k^2\,A^\dagger(k)A(k)
+ik\,A^\dagger(k)\Lambda(k)
-ik\,\Lambda^\dagger(k)A(k)\Bigr]\,.
\label{2.37}
\end{equation}

\bigskip
The quantization leads to

\begin{eqnarray}
\bigl[A(k),\,A^\dagger(p)\bigr]&=&0
=\bigl[A(k),\,A(p)\bigr]=\bigl[A^\dagger(k),\,A^\dagger(p)\bigr]\,,
\nonumber\\
\bigl[\Lambda(k),\,\Lambda^\dagger(p)\bigr]&=&-2\,k\,\delta(k-p)\,,
\nonumber\\
\bigl[\Lambda(k),\,\Lambda(p)\bigr]&=&0
=\bigl[\Lambda^\dagger(k),\,\Lambda^\dagger(p)\bigr]\,,
\nonumber\\
\bigl[A(k),\,\Lambda^\dagger(p)\bigr]&=&i\,\delta(k-p)\,,
\nonumber\\
\bigl[A(k),\,\Lambda(p)\bigr]&=&0
=\bigl[A^\dagger(k),\,\Lambda^\dagger(p)\bigr]\,.
\label{2.36}
\end{eqnarray}

\bigskip
We can verify that $A$ and $A^\dagger$ are raising and lowering
operators, respectively (whereas $\Lambda^\dagger$ and $\Lambda$ are
mixing ones).  However, we cannot talk on the spectrum of the field
$\phi$ because eigenvalues of $H_c$ are not necessarily positive and,
consequently, this does not permit us to introduce a vacuum state. At
this stage, it may be opportune to mention the paper of
ref.~\cite{Gi3}, where ghost fields have been conveniently introduced
to solve this problem and it was then possible to define a vacuum
state, but it was shown that this vacuum state is the only physical
state of the theory.

\bigskip
\section{Improved version of chiral-boson with linear constraints}
\bigskip
We now refer to the theory described by Lagrangian~(\ref{1.3}). Its
canonical momenta are the same as the Srivastava's model but we have
now three (continuous) constraints

\begin{eqnarray}
\Omega_1&=&q\,,
\label{3.1a}\\
\Omega_2&=&\pi-\phi^\prime\,,
\label{3.1b}\\
\Omega_3&=&\lambda^\prime\,.
\label{3.1c}
\end{eqnarray}

\bigskip\noindent
They satisfy the algebra (only nonvanishing brackets)

\begin{eqnarray}
\bigl\{\Omega_1(x,t),\,\Omega_3(y,t)\bigr\}&=&
\delta^\prime(x-y)\,,
\nonumber\\
\bigl\{\Omega_2(x,t),\,\Omega_2(y,t)\bigr\}&=&
-\,2\,\delta^\prime(x-y)\,.
\label{3.2}
\end{eqnarray}

\bigskip\noindent
It is opportune to observe that this theory, contrarily to what
occurs with the previous one, has just one physical degree of
freedom, the same number of the FJ theory.

\medskip
Considering Fourier transformations of fields and constraints, we have

\begin{eqnarray}
\tilde\Omega_1(k,t)&=&\tilde q(k,t)\,,
\label{3.3a}\\
\tilde\Omega_2(k,t)&=&\tilde\pi(k,t)-i\,k\,\tilde\phi(k,t)\,,
\label{3.3b}\\
\tilde\Omega_3(k,t)&=&i\,k\,\tilde\lambda(k,t)\,.
\label{3.3c}
\end{eqnarray}

\bigskip\noindent
The corresponding matrix $C(k,p)$ is

\begin{equation}
C(k,p)=\left(
\begin{array}{ccc}
0&0&ik\\
0&-2ik&0\\
ik&0&0
\end{array}
\right)
\,\delta(k+p)\,.
\label{3.4}
\end{equation}

\bigskip\noindent
As one observes, $\det C=0$ for $k=0$. Constraints $\tilde\Omega_1$
and $\tilde\Omega_2$ are first-class at this point (constraint
$\tilde\Omega_3$ is just the identity $0=0$). In order to quantize
the theory for all $k$, we have to fix the gauge. Natural choices are

\begin{eqnarray}
\tilde\lambda(0,t)&=&0\,,
\label{3.5a}\\
\tilde\phi(0,t)&=&0\,.
\label{3.5b}
\end{eqnarray}

\bigskip\noindent
Thus, the full set of constraints we have is

\begin{eqnarray}
\tilde\Omega_1(\bar k,t)&=&\tilde q(\bar k,t)\,,
\nonumber\\
\tilde\Omega_2(\bar k,t)&=&\tilde\pi(\bar k,t)
-i\,\bar k\,\tilde\phi(\bar k,t)\,,
\nonumber\\
\tilde\Omega_3(\bar k,t)&=&i\,\bar k\,\tilde\lambda(\bar k,t)\,,
\nonumber\\
\tilde\Omega_4&=&\tilde q(0,t)\,,
\nonumber\\
\tilde\Omega_5&=&\tilde\pi(0,t)\,,
\nonumber\\
\tilde\Omega_6&=&\tilde\lambda(0,t)\,,
\nonumber\\
\tilde\Omega_7&=&\tilde\phi(0,t)\,.
\label{3.6}
\end{eqnarray}

\bigskip\noindent
The constraint matrix is then enlarged to

\begin{equation}
C(k,p)=\left(
\begin{array}{ccccccc}
0&0&ik\delta(\bar k+\bar p)&0&0&0&0\\
0&-2i\delta(\bar k+\bar p)&0&0&0&0&0\\
i\delta(\bar k+\bar p)&0&0&0&0&0&0\\
0&0&0&0&0&-\delta(0)&0\\
0&0&0&0&0&0&-\delta(0)\\
0&0&0&\delta(0)&0&0&0\\
0&0&0&0&\delta(0)&0&0\\
\end{array}
\right)\,,
\label{3.7}
\end{equation}

\bigskip
that is not singular. The Dirac brackets are directly calculated

\begin{eqnarray}
\bigl\{\tilde\phi(\bar k,t),\,
\tilde\phi(\bar p,t)\bigr\}_D&=&
\frac{i}{2\bar k}\,\delta(\bar k+\bar p)\,,
\nonumber\\
\bigl\{\tilde\lambda(\bar k,t),\,
\tilde\lambda(\bar p,t)\bigr\}_D&=&0
=\bigl\{\tilde\phi(\bar k,t),\,
\tilde\lambda(\bar p,t)\bigr\}_D\,.
\label{3.8}
\end{eqnarray}

\bigskip\noindent
Other brackets involving $\tilde\phi(0)$ and $\tilde\lambda(0)$ are
also zero. We notice here that mixing brackets involving $\lambda$
and $\phi$ are zero and that the first bracket above is the same as
the FJ theory.

\medskip
To obtain the spectrum we have to calculate the canonical
Hamiltonian. Taken into account the same comments made at Sec.~3, we
get

\begin{eqnarray}
H_c&=&\int dx\,\bigl(\phi^\prime\phi^\prime
+\lambda^\prime\phi\bigr)\,,
\nonumber\\
&\rightarrow&\int dx\,\phi^\prime\phi^\prime\,.
\label{3.9}
\end{eqnarray}

\bigskip\noindent
In the last step it was used the constraint $\Omega_3$. As this is
also the same canonical Hamiltonian of the FJ theory, it is not
difficult to see that the field $\phi$ has the same spectrum of the
FJ theory and that $\lambda$ does not generate any state. We can thus
conclude that ${\cal L}_1$ and ${\cal L}_3$ describe the same theory,
classically and quantically, contrarily to what occurs with  ${\cal
L}_1$ and ${\cal L}_2$ that are equivalently just in terms of
equation of motion.

\section{Conclusion}

We have made a comparative study among chiral-boson theories, mainly
in their quantum aspects. In the case of the FJ formulation there is
a zero mode where the constraint becomes first-class~\cite{Gi1}. We
have chosen a convenient gauge-fixing in order to define creation and
destruction operators at all points of the momentum space.
Concerning chiral-boson with a linear constraint, we have also
reviewed the obtainment of the spectrum and discuss the equivalence
with the FJ theory. Here, constraints are second-class at all points
but it is not possible to define a spectrum~\cite{Gi2}.

\medskip
The main part of our work was a discussion of a improved version of
the chiral boson with linear constraint. We have shown that it has
the same physical degrees of freedom of the FJ theory and also
exhibits a point where constraints are first-class. We have concluded
by carefully studying its spectrum and showing that it is also the
same of the FJ theory.

\bigskip
\noindent {\bf Acknowledgment:} This work is supported in part by
Conselho Nacional de Desenvolvimento Cient\'{\i}fico e Tecnol\'ogico
- CNPq (Brazilian Research Agency).

\end{document}